\documentclass[twocolumn,twoside,slac_two]{revtex4}
\usepackage{graphicx}
\usepackage{fancyhdr}
\usepackage{mkPhys}

\renewcommand{\babar}{\mbox{\sl BABAR}\xspace}

\begin{document}

\title{Direct \CP violation in $B$ decays}

%

\author{M. Kreps}
\affiliation{Institut f\"ur Experimentelle Kernphysik, University of Karlsruhe, Germany}

\begin{abstract}
  We review recent experimental results on direct \CP
violation. The hot topic is a measurement in charmless two-body
decays of $B^0$, $B^+$. In connection to this the first
analogous measurements in $B_s^0$ and $\Lambda_b^0$ decays
are now available. Furthermore first evidence for direct \CP
violation in $B^+$ decays is obtained from Dalitz plot
analyzes of the $K^+\pi^-\pi^+$ final state at B-factories. The
last group of discussed results probes the $b\rightarrow
c\bar{c}d$ transition in attempt to resolve the discrepancy
between Belle and \babar experiments in \CP violation in the 
$B^0\rightarrow D^+D^-$ decays.
\end{abstract}

\maketitle

\thispagestyle{fancy}


\section{Introduction}
Measurements of the direct \CP violation form an important
test of the CKM mechanism of \CP violation in the standard model.
In addition it provides a window for searches for new
physics beyond the standard model. 
Direct \CP violation is a decay property, where the
amplitudes for the processes $B\rightarrow f$
and $\overline{B}\rightarrow \overline{f}$ are different.
Experimentally it manifests as the difference in the decay
widths of the two charge conjugated states. In the case of
decays to the flavor
eigenstates, the experimental observable is the time integrated
asymmetry
      \begin{eqnarray}
         A_{\CP}\,=\,\frac{\Gamma(B\rightarrow f)-\Gamma(\overline{B}\rightarrow \overline{f})}
                    {\Gamma(B\rightarrow f)+\Gamma(\overline{B}\rightarrow \overline{f})}.
  \nonumber
      \end{eqnarray}
For decays to a common final state one has to disentangle the
\CP violation induced by the mixing from the direct \CP violation.
This requires a study of the time dependent asymmetry 
      \begin{eqnarray}
        A_{\CP}(t)&=&\frac{\dd\Gamma/\dd t(\overline{B}\rightarrow f)-\dd\Gamma/\dd t(B\rightarrow f)}
                        {\dd\Gamma/\dd t(\overline{B}\rightarrow f)+\dd\Gamma/\dd t(B\rightarrow f)} \nonumber \\
              &=&
                   \mathcal{S}\sin{\Delta mt}+\mathcal{A}\cos{\Delta mt},
       \nonumber
      \end{eqnarray}
where $\Delta m$ is the mixing frequency, $\mathcal{S}$ is the \CP
violation in the interference of the decays with and without
mixing, and $\mathcal{A}$ is the direct \CP violation.

To have an observable direct \CP violation, at least two
interfering amplitudes with different weak and strong phases
are required. In case of the two amplitudes $A_1$ and $A_2$
with relative weak phase $\phi$ and relative strong phase
$\delta$ the decay widths are
\begin{eqnarray}
  \Gamma(B\rightarrow f)&\propto&
|A_1+A_2\mathrm{e}^{i\phi}\mathrm{e}^{i\delta}|^2 \nonumber \\
\Gamma(\overline{B}\rightarrow
\overline{f})&\propto&|A_1+A_2\mathrm{e}^{-i\phi}\mathrm{e}^{i\delta}|^2.
\nonumber
\end{eqnarray}
With some algebra one can easily see that the asymmetry is
\begin{eqnarray}
  A_{\CP}\propto \sin\phi\sin\delta.
  \nonumber
\end{eqnarray}
The necessary condition for an observable direct
\CP violation is to have both the relative weak phase as
well as the relative strong phase different from $0$ or $\pi$.

The direct \CP violation is rather well understood
theoretically, but difficult to predict. The difficulty
comes from the fact that  $A_{\CP}$ depends not only on
the weak phase, but also on the strong phase, which involves
non-perturbative long distance effects and it is this part,
which makes predictions difficult.

The importance of the direct \CP violation is manifold. It is
crucial for determining the angle $\gamma$ of the
Cabibbo-Kobayashi-Maskawa (CKM) matrix. In addition it can 
provide useful tests of the theoretical tools as well as
tests of the CKM mechanism of \CP violation. It also has 
potential for the discovery of physics beyond the standard model.
The best decays for searches of new physics are those,
which are dominated by a single CKM phase. Observation of
the direct \CP violation in such decays would imply the presence of
a second amplitude and thus evidence for new physics.

In this paper we review recent experimental results on
direct \CP violation. We start in section \ref{sec:2} with
the charmless two-body decays of $B^0$, $B^+$, $B_s^0$, and
$\Lambda_b^0$.  In section \ref{sec:3} we discuss the first evidence for direct
\CP violation in $B^+\rightarrow K^+\pi^-\pi^+$ decays.
Section \ref{sec:4} reviews measurements in the $b\rightarrow
c\bar{c}d$ transition with section \ref{sec:5} focusing on
\BplusJpsiK decays.

\section{Charmless two body $b$-hadron decays}
\label{sec:2}

Direct \CP violation is most thoroughly studied in
$b$-hadron decays to charmless two body final states.
Several dozens of
different decays are already studied experimentally. 
\begin{figure}[bt]
\begin{center}
 \includegraphics[width=7.0cm]{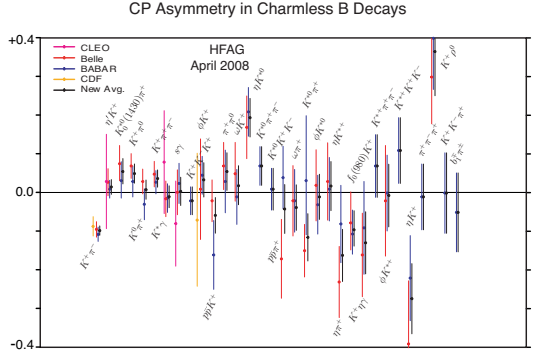}
 \caption{Summary of the most precise measurements of direct \CP
violation in charmless two body decays of the $B^+$ and
$B^0$ mesons to a flavor specific final state \cite{Barberio:2007cr}.
\label{fig1}}
\end{center}
\end{figure}
In Fig.~\ref{fig1} we show a summary of the most precise
measurements in decays to a flavor specific final state \cite{Barberio:2007cr}.

\subsection{$B^0\rightarrow K^+\pi^-$ and $B^+\rightarrow
K^+\pi^0$ decays}
The  \CP violation in the decay $B^0\rightarrow K^+\pi^-$ is
the only direct \CP violation, which has been observed in a single experiment with more than 5
standard deviations. Together with results in the decay
$B^+\rightarrow K^+\pi^0$ this generated a considerable amount of
interest. In the case of $B^+\rightarrow K^+\pi^0$, two
measurements are available. The Belle experiment \cite{Lin:2008} used $535$
million $B\overline{B}$ pairs and measures 
\begin{eqnarray}
A_{\CP}=0.07\pm 0.03\pm 0.01.  \nonumber
\end{eqnarray}
The \babar experiment \cite{Aubert:2004aq}, using $383$ million
$B\overline{B}$ pairs,  measures in the decay $B^+\rightarrow
K^+\pi^0$
\begin{eqnarray}
A_{\CP}=0.03\pm0.04\pm0.01  .\nonumber
\end{eqnarray}
Averaging those two measurements yields
\begin{eqnarray} 
A_{\CP}=0.05\pm0.025. \nonumber
\end{eqnarray}
Both  experiments use thier sample also to measure the direct
\CP asymmetry in the $B^0\rightarrow K^+\pi^-$ decay.
Belle obtains \cite{Lin:2008}
\begin{eqnarray}
A_{\CP}=-0.094\pm0.018\pm0.008 \nonumber
\end{eqnarray}
with a significance of $4.8$ standard deviations. \babar
measures \cite{Aubert:2007mj}
\begin{eqnarray}
A_{\CP}=-0.107\pm0.018^{+0.007}_{-0.004} \nonumber
\end{eqnarray}
with a significance of $5.5$ standard deviations. Two other
measurements exist for the $B^0\rightarrow K^+\pi^-$ decay.
The CLEO experiment measures
\cite{Chen:2000hv}
\begin{eqnarray}
A_{\CP}=-0.040\pm0.160\pm0.020 \nonumber
\end{eqnarray}
and the CDF experiment, using $1$ \invfb of
$p\overline{p}$ collisions, measures
\cite{CDF_B0_public_note}
\begin{eqnarray}
A_{\CP}=-0.086\pm0.023\pm0.009, \nonumber
\end{eqnarray}
which has a significance of $3.5$ standard deviations.
Averaging the four mentioned measurements one obtains
\begin{eqnarray}
A_{\CP}=-0.097\pm0.012. \nonumber
\end{eqnarray}

\begin{figure}[htb]
\begin{center}
 \includegraphics[width=7.0cm]{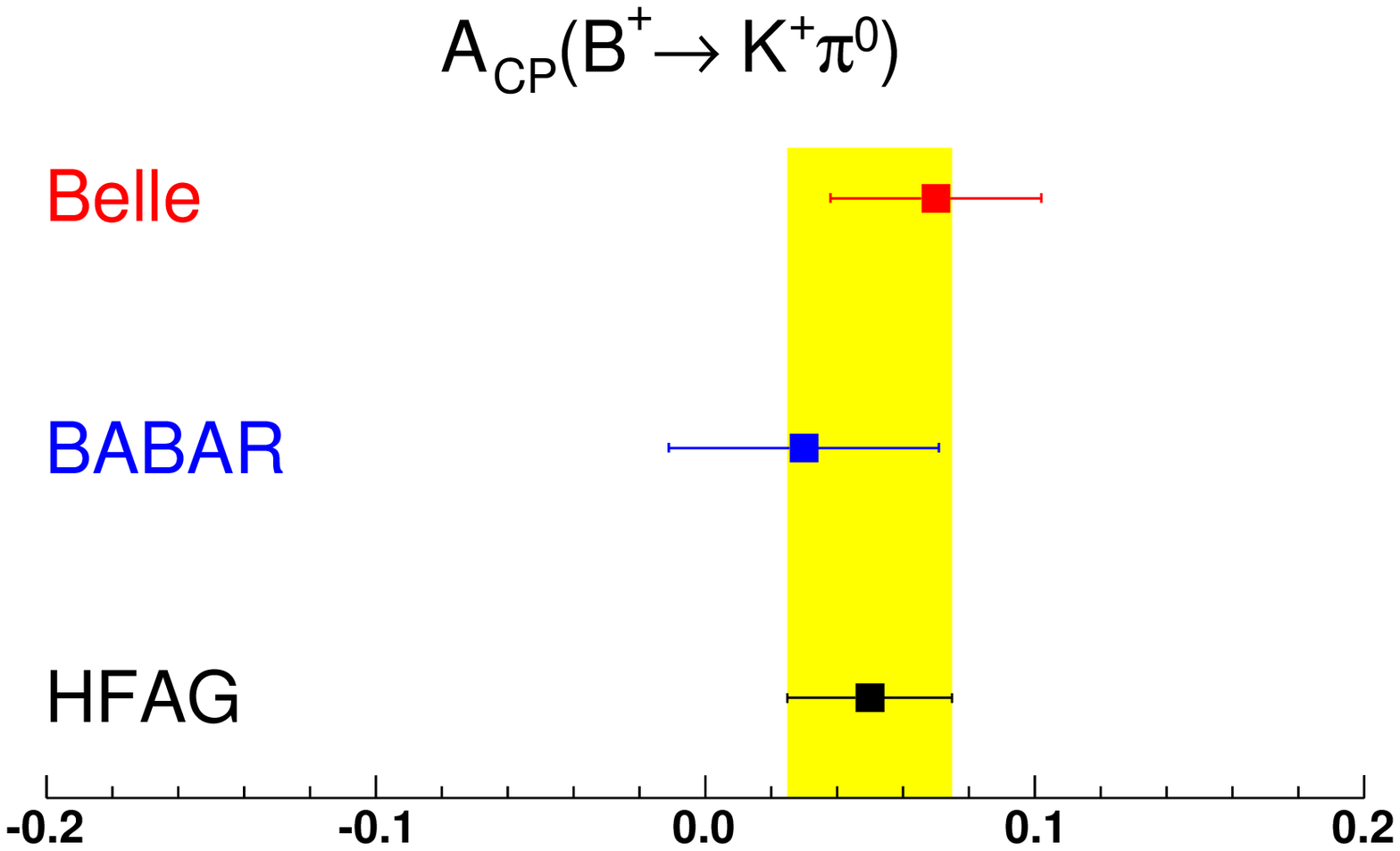}
 \includegraphics[width=7.0cm]{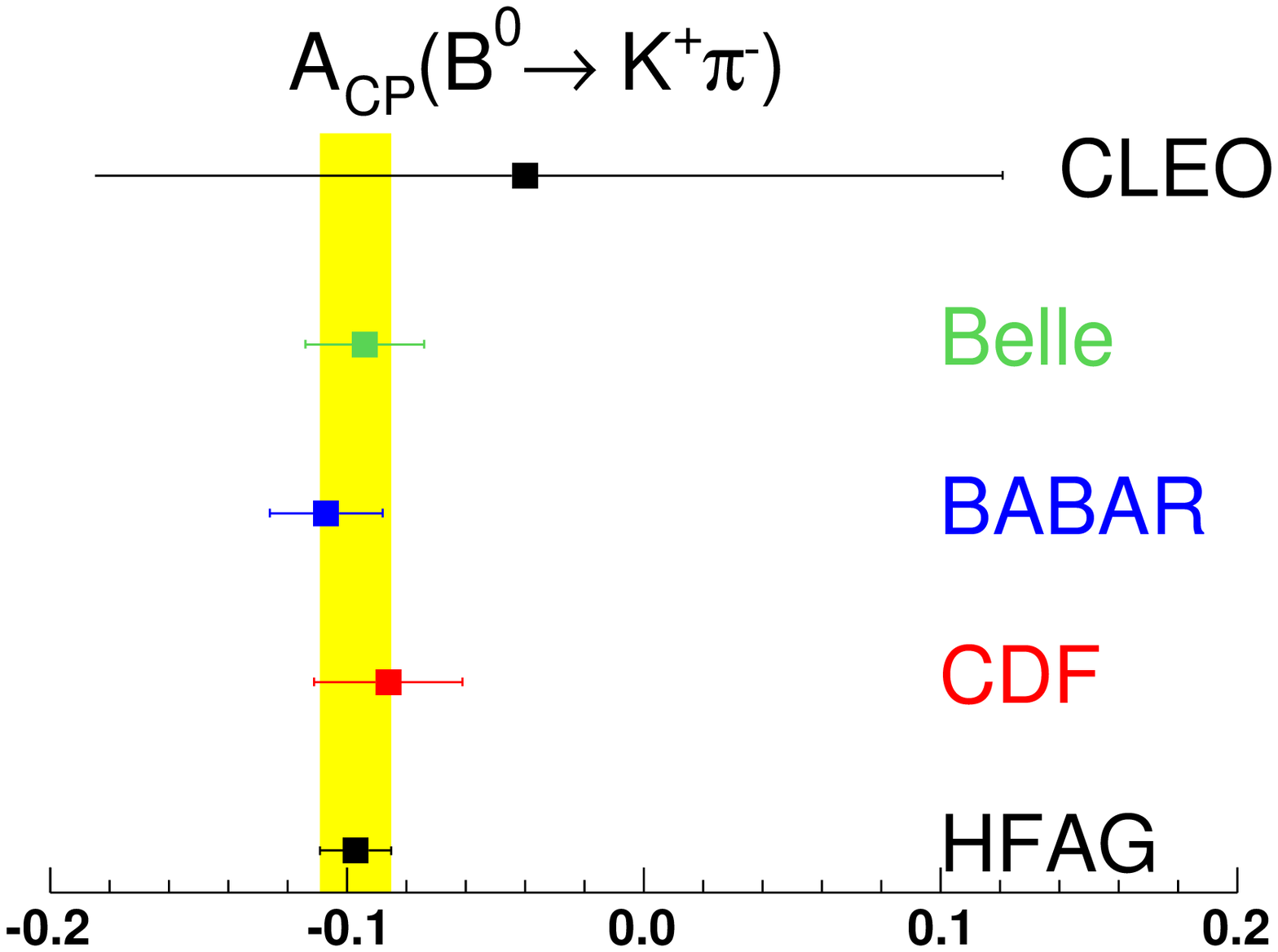}
 \caption{Summary of the measurements of the direct \CP
violation in the $B^0\rightarrow K^+\pi^-$ (bottom) and the $B^+\rightarrow
K^+\pi^0$ (top) decays.
\label{fig2}}
\end{center}
\end{figure}
In Fig.~\ref{fig2} we show a graphical summary of the
measurements. In both decays, all measurements are
consistent with each other. What is clearly different is
the direct \CP asymmetry in $B^0$ and $B^+$ decays. This
difference is in contradiction with the na\"\i ve expectation
that both asymmetries should be the same. 
\begin{figure}[htb]
\begin{center}
 \includegraphics[angle=90,width=8.0cm]{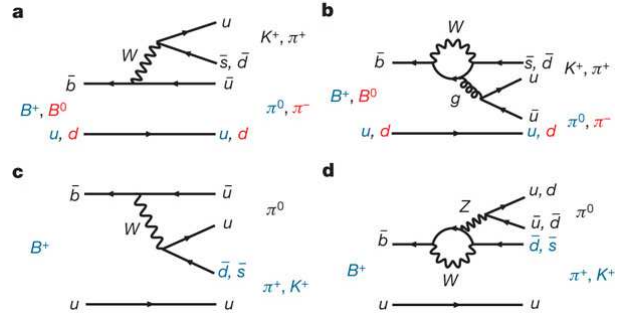}
 \caption{Contributions to the decay $B\rightarrow K\pi$.
The possible processes are a) tree, b) penguin, c)
color suppressed tree, and d) electroweak penguin amplitudes.
The first two contribute to both $B^0\rightarrow K^+\pi^-$ and
$B^+\rightarrow K^+\pi^0$ decays, while the last two contribute
only to the decay $B^+\rightarrow K^+\pi^0$. The figure is
reproduced from Ref.~\cite{Lin:2008}.
\label{fig3}}
\end{center}
\end{figure}
In Fig.~\ref{fig3} Feynman diagrams of all contributions to
the given decays are
shown. The contributions are tree (Fig.~\ref{fig3}a), penguin
(Fig.~\ref{fig3}b), color suppressed tree
(Fig.~\ref{fig3}c),
and electroweak penguin (Fig.~\ref{fig3}d) amplitudes. While
the first two
are present in both $B^+$ and $B^0$ decays, the other two are
present only in the $B^+$ decays. The na\"\i ve expectation that the
two direct \CP asymmetries should be same comes from
the neglection of color suppressed tree and electroweak
penguin amplitudes in the case of $B^+$ decays. While the difference
in \CP asymmetry could be generated by a new physics
contribution to the electroweak penguin amplitude
\cite{Hou:2006jy, Peskin:2008}, it
is also too early to exclude that the standard model with
enhanced color suppressed tree amplitude can explain
the observed difference \cite{Gronau:2007jm}.
It can be concluded that while the situation is tantalizing, it is not
conclusive and it is too early to claim the observation of  new
physics beyond the standard model.

\subsection{$B_s\rightarrow K^-\pi^+$ decays}

While at the B-factories 
one can study only $B^+$ and $B^0$ mesons, the Tevatron
experiments have access to all $b$-hadron species.
One of the important outcomes is that the CDF experiment,
together with the direct \CP violation in
$B^0\rightarrow K^+\pi^-$, also measures the direct \CP
violation in the decay
$B_s\rightarrow K^-\pi^+$ \cite{CDF_B0_public_note}.

The standard model expectations vary strongly with
the calculation method and generally have large uncertainties
\cite{Beneke:2003zv,Ali:2007ff,Williamson:2006hb}.
In the context of the standard model a relation between the amplitudes
in $B^0\rightarrow K^+\pi^-$ and $B_s\rightarrow K^-\pi^+$
decays was predicted
\cite{He:1998rq,Deshpande:2000jp,Gronau:2000md,Gronau:2000zy,Lipkin:2005pb}. 
In terms of the decay widths it has the form
\begin{eqnarray}
\frac{\Gamma(\overline{B}^0\rightarrow
K^-\pi^+)-\Gamma(B^0\rightarrow K^+\pi^-)}{
\Gamma(B_s^0\rightarrow
K^-\pi^+)-\Gamma(\overline{B}_s^0\rightarrow K^+\pi^-)}=1.
\nonumber
\end{eqnarray}
This relation can be translated into a relation between the direct
\CP asymmetries between $B^0\rightarrow K^+\pi^-$ and
$B_s\rightarrow K^-\pi^+$ decays and has the form
\begin{eqnarray}
  \frac{A_{\CP}(B_s^0\rightarrow K^-\pi^+)}
     {A_{\CP}(B^0\rightarrow K^+\pi^-)}=-
  \frac{\mathcal{B}(B^0\rightarrow K^+\pi^-)\tau(B^0)}
   {\mathcal{B}(B_s^0\rightarrow K^-\pi^+)\tau(B_s)},
  \nonumber
\end{eqnarray}
where \Br denotes the corresponding branching fraction and
$\tau$ the mean lifetime. Using measured values for \Br,
lifetimes, and $A_{\CP}(B^0\rightarrow K^+\pi^-)$
\cite{Barberio:2007cr} one
obtains 
\begin{equation}
A_{\CP}(B_s^0\rightarrow K^-\pi^+)\approx +0.37.
\label{eq9}
\end{equation}

\begin{figure}[tb]
\begin{center}
 \includegraphics[width=7.0cm]{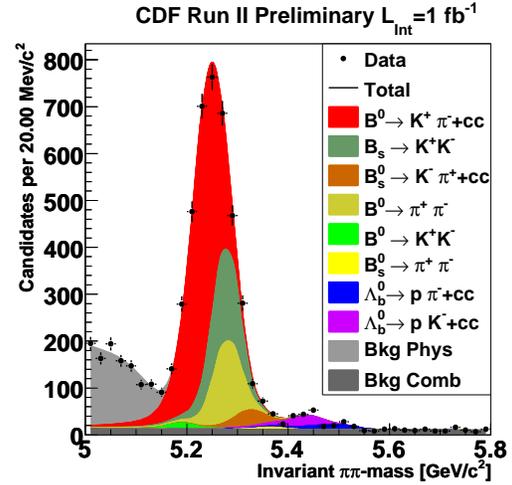}
 \caption{Typical invariant mass distribution in the analysis of charmless two
body $b$-hadron decays at CDF
\cite{CDF_Lambdab_public_note}. 
\label{fig4}}
\end{center}
\end{figure}
In Fig.~\ref{fig4} we show a typical mass distribution
observed by CDF using $1$ \invfb of data in the charmless two body
$b$-hadron decays \cite{CDF_Lambdab_public_note}. 
The signal for the decay
$B_s^0\rightarrow K^-\pi^+$ can be clearly separated and CDF
measures
\begin{eqnarray}
  A_{\CP}(B_s^0\rightarrow K^-\pi^+)   = +0.39\pm0.15\pm0.08. 
\end{eqnarray}
The measurement is well consistent with the expectation
from equation~\ref{eq9}. It has $2.5$ standard deviations significance of
being nonzero. As CDF
already has about $3$ \invfb of data
available for analysis, it will be interesting to watch out
for an analysis update, which has the potential to provide the first evidence for
direct \CP violation in the $B_s^0$ system.

\subsection{$\Lambda_b^0\rightarrow pK^-$ and
$\Lambda_b^0\rightarrow p\pi^-$ decays}

\begin{figure}[tb]
\begin{center}
 \includegraphics[width=7.0cm]{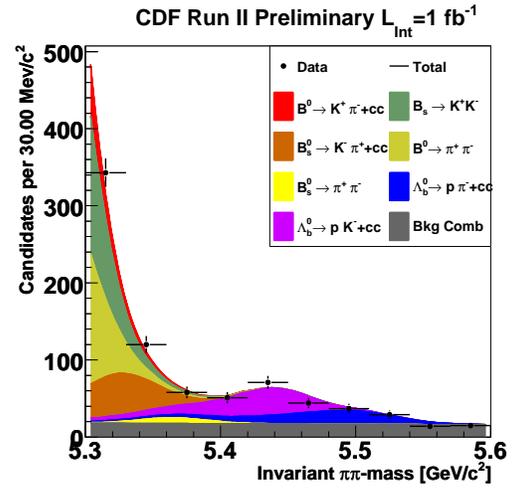}
 \caption{Invariant mass distribution of charmless two body
decays. Shown is a more detailed view from Fig.~\ref{fig4},
focusing on the region
with $\Lambda_b^0$ decays \cite{CDF_Lambdab_public_note}.
\label{fig5}}
\end{center}
\end{figure}

Two more decays in Fig.~\ref{fig4} are worth of mentioning.
Those are the decays $\Lambda_b^0\rightarrow pK^-$ and
$\Lambda_b^0\rightarrow p\pi^-$. In Fig.~\ref{fig5} we show
a zoom of Fig.~\ref{fig4} into the mass region from $5.3$ to
$5.6$ \gevcc with a clear $\Lambda_b^0$ signal.
This allows for the first time to study the direct \CP
violation in $b$-baryon decays in experiment.

The decays of $\Lambda_b^0$ to $pK^-$ and $p\pi^-$ are
not very well studied theoretically. Mohanta {\it et al.} \cite{Mohanta:2000nk}
predict that in the standard model the \CP asymmetries can
reach a size of the order of 10\%. On the other hand in
supersymmetry models with R-parity violation the direct \CP
violation could be significantly suppressed \cite{Mohanta:2000za}.

With the sample shown in Fig.~\ref{fig5}, CDF  measures
the following \CP asymmetries
\begin{eqnarray}
      A_{\CP}(\Lambda_b^0\rightarrow
p\pi)&=&0.03\pm0.17\pm0.05, \nonumber \\
      A_{\CP}(\Lambda_b^0\rightarrow
pK)&=&0.37\pm0.17\pm0.03. \nonumber
\end{eqnarray}
Both decays are still compatible with no direct \CP
violation, but in the case of $\Lambda_b^0\rightarrow pK^-$ it is
$2.1$ standard deviations away from zero. Also both results
are compatible with the expected value of
direct \CP violation. We can expect that at least in
case of the decay $\Lambda_b^0\rightarrow pK^-$ the next round of
analysis by CDF will provide first evidence for direct \CP
violation in the $b$-baryon sector. Until then it would be
welcome if the theory expectations could be refined for those decays.

\section{Dalitz plot analysis of $B^+\rightarrow K^+\pi^-\pi^+$}
\label{sec:3}

Many of the measurements shown in Fig.~\ref{fig1} involve
broad resonances. The experimental study of those requires
Dalitz plot analyses to resolve the broad resonances,
which are often overlapping. Several three body final states
were studied by Belle and \babar using the Dalitz plot technique. The recent results
include $B^0\rightarrow K^+\pi^-\pi^0$
\cite{Chang:2004um,Aubert:2007bs}, $B^0\rightarrow
K^+K^-K^0$ \cite{Aubert:2007sd}, $B^0\rightarrow
K_s\pi^+\pi^-$ \cite{Aubert:2007vi} and most importantly $B^+\rightarrow
K^+\pi^-\pi^+$ \cite{Garmash:2005rv,Aubert:2008bj}, 
which we discuss here in some detail. 

\begin{figure}[tb]
\begin{center}
 \includegraphics[width=7.0cm]{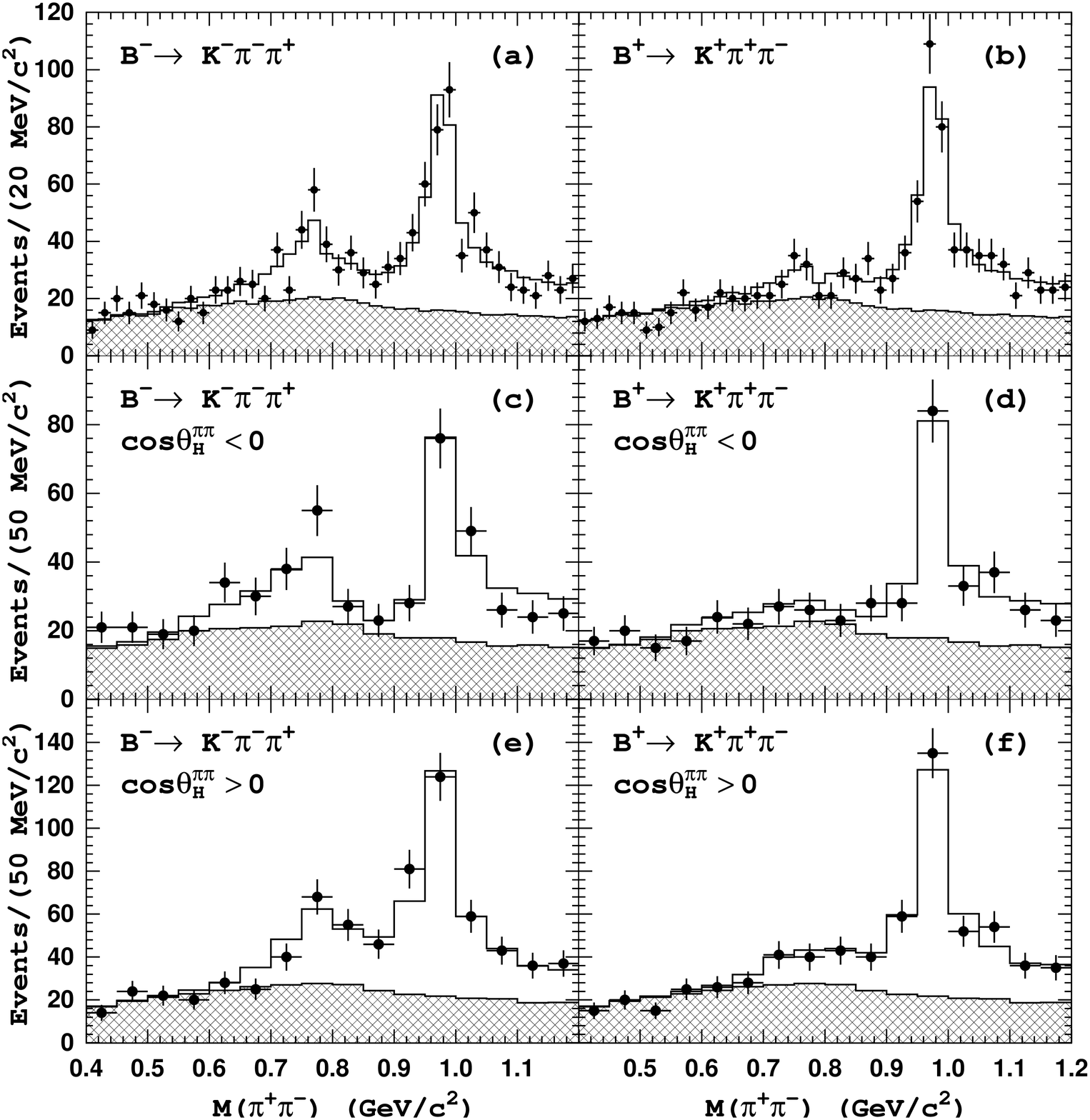}
 \caption{Projection of the Dalitz plot fit from Belle of
$B^+\rightarrow K^+\pi^-\pi^+$ on the $\pi^+\pi^-$ invariant
mass in the region of the $\rho^0(770)$ and $f_0^0(980)$ resonances
\cite{Garmash:2005rv}. The left column
shows $B^-$ while the right one shows $B^+$. 
The top row contains all
events. The middle and bottom row shows two distinct
kinematic regions.
\label{fig6}}
\end{center} 
\end{figure}

\begin{figure}[tb]
\begin{center}
 \includegraphics[width=7.0cm]{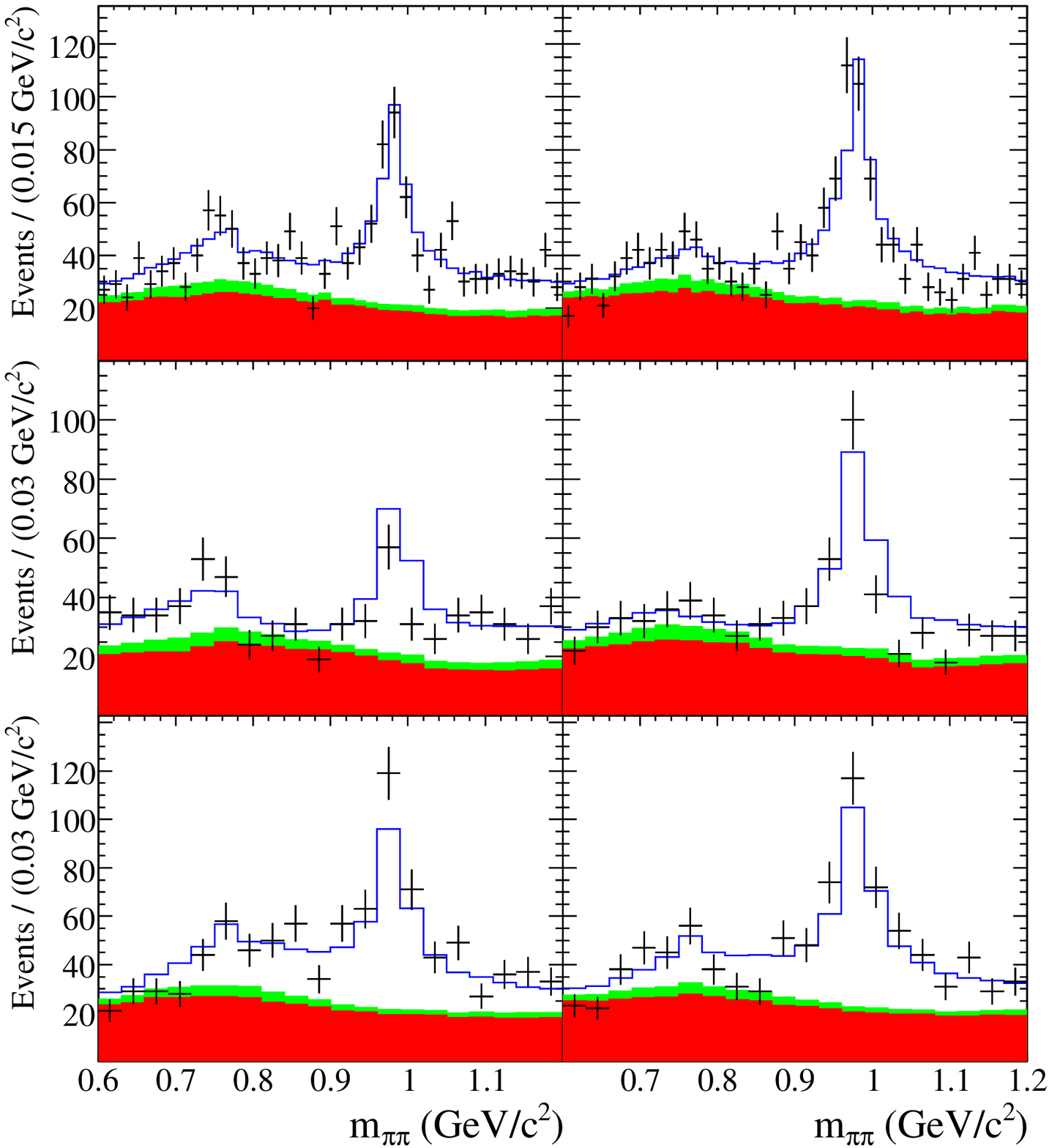}
 \caption{Projection of the Dalitz plot fit from \babar of
$B^+\rightarrow K^+\pi^-\pi^+$ on the $\pi^+\pi^-$ invariant
mass in the region of the $\rho^0(770)$ and $f_0^0(980)$ resonances
\cite{Aubert:2008bj}. The left column
shows $B^-$ while the right one shows $B^+$. 
The top row contains all
events. The middle and bottom row shows two distinct
kinematic regions.
\label{fig7}}
\end{center} 
\end{figure}

The Dalitz plot analyses of $B^+\rightarrow K^+\pi^-\pi^+$
are performed by both Belle and \babar 
using $\approx 380$ million $B\overline{B}$ pairs. The
chosen Dalitz model slightly
differs between the two experiments, but the extracted
branching fractions for different components are consistent.
The total \CP asymmetry of the
decay $B^+\rightarrow K^+\pi^-\pi^+$ over the full Dalitz space is measured by Belle
 to be 
\begin{eqnarray}
A_{\CP}(B^+\rightarrow K^+\pi^-\pi^+)=0.049\pm0.026
\pm0.020. \nonumber
\end{eqnarray}
\babar measures the same quantity to be
\begin{eqnarray}
A_{\CP}(B^+\rightarrow
K^+\pi^-\pi^+)=0.028\pm0.02\pm0.02\pm0.012. \nonumber
\end{eqnarray}
The last uncertainty is due to the Dalitz model, which
in case of the Belle result is included in the systematic
uncertainty. In addition to the measurement over the full Dalitz
plot, also measurements in different quasi-two-body final
states were performed. In Fig.~\ref{fig6} we show the projection
of the Dalitz plot fit from Belle on the invariant mass of
the two pions,
separately for $B^+$ and $B^-$. The analogous distributions from
\babar are shown in Fig.~\ref{fig7}.
The two decays which are noteworthy are $B^+\rightarrow
\rho^0(770) K^+$ and $B^+\rightarrow f_2^0(1280)K^+$.
For those Belle  measures
\begin{eqnarray}
  A_{\CP}(B^+\rightarrow\rho^0(770) K^+) &=&
0.30\pm0.11^{+0.11}_{-0.05}, \nonumber \\
  A_{\CP}(B^+\rightarrow f_2^0(1280)K^+) &=&
-0.59\pm0.22\pm0.04,  \nonumber
\end{eqnarray}
and \babar obtains
\begin{eqnarray}
  A_{\CP}(B^+\rightarrow\rho^0(770) K^+) &=&
0.44\pm0.10^{+0.06}_{-0.14},  \nonumber \\
  A_{\CP}(B^+\rightarrow f_2^0(1280)K^+) &=& -0.85\pm0.22^{+0.26}_{-0.13}.  \nonumber
\end{eqnarray}
The asymmetry $A_{\CP}(B^+\rightarrow\rho^0(770) K^+)$ has
a significance of $3.9$ and $3.7$ standard deviations for Belle
and \babar respectively, thus providing first evidence for
the direct \CP violation in $B^+$ decays. Additionally, the
asymmetry $A_{\CP}(B^+\rightarrow f_2^0(1280)K^+)$ from \babar shows
a significance of more than $3$ standard deviations, but this
decreases below $3$ standard deviations with some variations
in the Dalitz
model. Direct \CP asymmetries of all other decay modes in
the Dalitz model are compatible with zero.

\section{$b\rightarrow c\bar{c}d$ transition}
\label{sec:4}

Another topic which recently got a considerable amount of attention
is the \CP violation in $b\rightarrow c\bar{c}d$
transitions. While the standard model expects tiny direct \CP
violation in decays, governed by the $b\rightarrow c\bar{c}d$ quark
level transition, the measurement in $B^0\rightarrow D^+D^-$
decays by Belle yielded an unusually high value of \cite{Fratina:2007zk}
\begin{eqnarray}
A_{\CP}(B^0\rightarrow D^+D^-)=-0.91\pm0.23\pm0.06. \nonumber
\end{eqnarray}
On the other hand the same measurement by \babar
 \cite{Aubert:2007pa} resulted in
\begin{eqnarray}
A_{\CP}(B^0\rightarrow D^+D^-)=0.11\pm0.22\pm0.07, \nonumber
\end{eqnarray}
which is consistent with expectations, but inconsistent with
the Belle measurement. This discrepancy and the fact that Belle
measured a large \CP violation prompted large interest and
more work on related decays governed by the $b\rightarrow
c\bar{c}d$ quark level transition.

\subsection{$B^+\rightarrow D^+\overline{D}^0$}

\begin{figure}[tb]
\begin{center}
 \includegraphics[width=7.0cm]{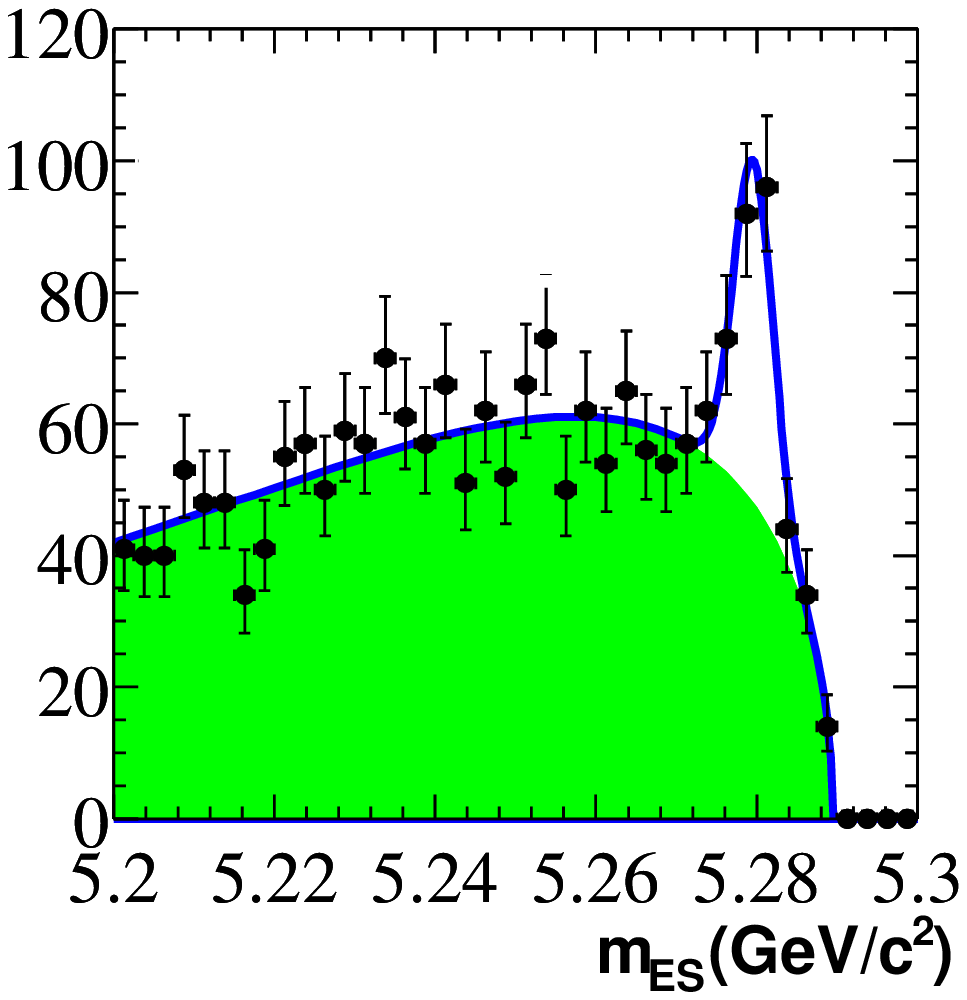}
 \caption{The beam constrained mass distribution of
the $B^+\rightarrow  D^+\overline{D}^0$ events from \babar
\cite{Aubert:2006ia}.
\label{fig8}}
\end{center} 
\end{figure}

The decay $B^+\rightarrow D^+\overline{D}^0$ is an analog of the
$B^0\rightarrow D^+D^-$ decay proceeding with the same quark
level transition. This decay was previously analyzed by
\babar using $231$ million $B\overline{B}$ pairs
\cite{Aubert:2006ia}. In Fig.~\ref{fig8} the beam
constrained mass distribution of $B^+\rightarrow
D^+\overline{D}^0$ events is shown. From $129\pm20$ signal events
\babar extracts
\begin{eqnarray}
A_{\CP}(B^+\rightarrow
D^+\overline{D}^0)=-0.13\pm0.14\pm0.02. \nonumber
\end{eqnarray}
\begin{figure}[tb]
\begin{center}
 \includegraphics[width=7.0cm]{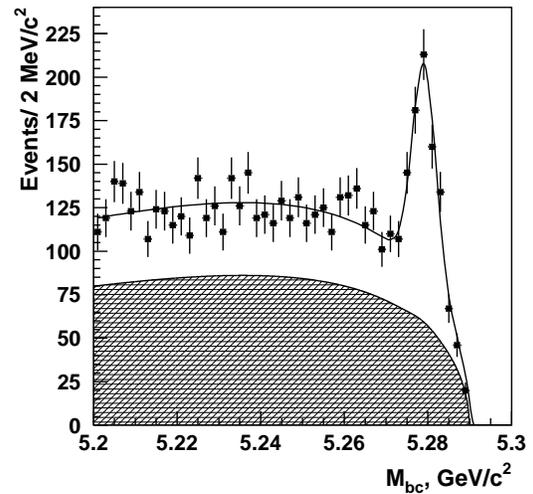}
 \caption{The beam constrained mass distribution of
the $B^+\rightarrow  D^+\overline{D}^0$ events from Belle 
\cite{Adachi:2008cj}.
\label{fig9}}
\end{center} 
\end{figure}
Recently Belle performed the same analysis using
$657$ million $B\overline{B}$ pairs \cite{Adachi:2008cj}.
The beam constrained mass distribution is shown in
Fig.~\ref{fig9}. Using $194.2\pm20.4$ signal events the
measured direct \CP asymmetry is
\begin{eqnarray}
A_{\CP}(B^+\rightarrow D^+\overline{D}^0)= 0.00\pm0.08\pm0.02.
\nonumber
\end{eqnarray}

Both measurements are consistent with no direct \CP
violation, as expected within the standard model. But it is also
difficult to make a firm statement about the \CP violation in the
decay $B^0\rightarrow D^+D^-$ as spectator effects are
different between $B^+\rightarrow  D^+\overline{D}^0$ and
$B^0\rightarrow D^+D^-$ decays so that the weak exchange and weak annihilation
diagrams are suppressed in the $B^+$ case compared to the
$B^0$ decay. While the two diagrams
are expected to be negligible in both $B^+$ and $B^0$ decays, 
it cannot be excluded that
$B^0$ and $B^+$ would have different direct \CP violation in
those decays. A recent theoretical discussion of the
$B\rightarrow D\overline{D}$ decays can be found in
Ref.~\cite{Gronau:2008ed}.

\subsection{$B^0\rightarrow D^{*+}D^{*-}$}

Another decay in which one can test the $b\rightarrow c\bar{c}d$
transition is the decay $B^0\rightarrow D^{*+}D^{*-}$. This
decay proceeds through the same diagrams as the decay $B^0\rightarrow
D^+D^-$ and is therefore very useful in trying to understand
the discrepancy between Belle and \babar in the $B^0\rightarrow
D^+D^-$ decay. There is one complication in the measurement as
the final state consists of two vector particles for which decays
through different orbital momenta are possible. This causes
that the final state is a mixture of \CP-even and \CP-odd final
states, which needs to be taken into account.

\begin{figure}[tb]
\begin{center}
 \includegraphics[width=7.0cm]{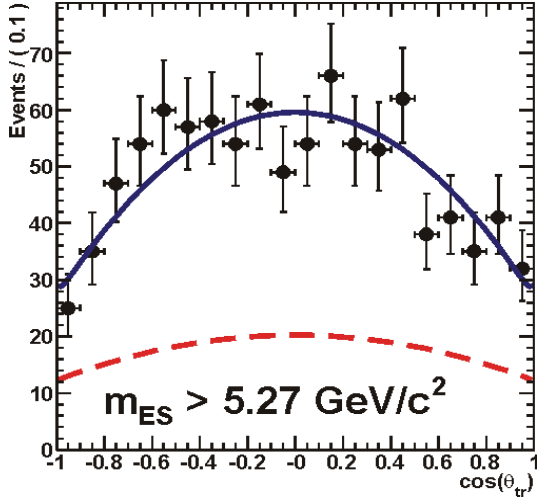}
 \caption{Distribution of $\cos\theta_{tr}$ used to
disentangle the \CP-even and \CP-odd fraction in  $B^0\rightarrow
D^{*+}D^{*-}$ from \babar \cite{Aubert:2007rr}.
The dashed line represents the background distribution, the full
line the fit projection, and the points with error bars show data.
\label{fig10}}
\end{center} 
\end{figure}
\begin{figure}[tb]
\begin{center}
 \includegraphics[width=7.0cm]{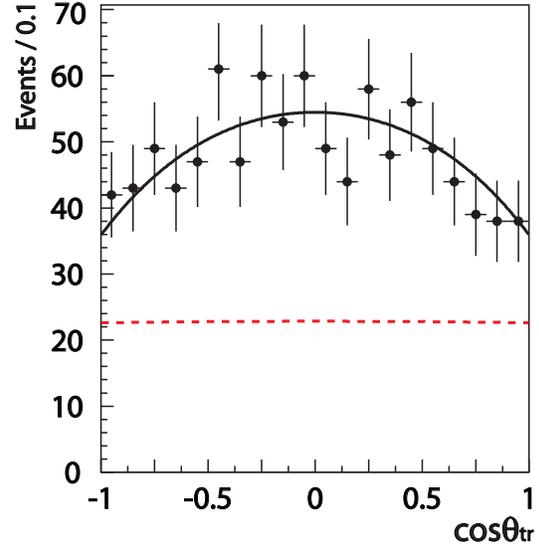}
 \caption{Distribution of  $\cos\theta_{tr}$ used to
disentangle the \CP-even and \CP-odd fraction in  $B^0\rightarrow
D^{*+}D^{*-}$ from Belle \cite{Aushev:2008}.
The dashed line represents the background distribution, the full
line the fit projection, and the points with error bars show data.
\label{fig11}}
\end{center} 
\end{figure}
\babar uses $383$ million $B\overline{B}$ pairs
providing a signal of $617\pm33$ $B^0\rightarrow D^{*+}D^{*-}$
decays \cite{Aubert:2007rr}. Belle recently presented 
a measurement based on $535$ million $B\overline{B}$ pairs,
which provides $545\pm29$ signal events \cite{Aushev:2008}. Both experiments
use a $\cos\theta_{tr}$ distribution to disentangle the \CP-even
and \CP-odd fraction on a statistical basis. $\theta_{tr}$
is the polar angle of the pion from the $D^{*+}$ decay in the $D^{*+}$ rest
frame with the $z$-axis being normal to the $D^{*-}$ decay plane and
the $x$-axis opposite to the $D^{*-}$ momentum.
The $\cos\theta_{tr}$ distribution together with the fit
projection is shown for \babar  in
Fig.~\ref{fig10} and for Belle  in
Fig.~\ref{fig11}.
The measured fraction of \CP-odd component is
$0.143\pm0.034\pm0.008$ for \babar and
$0.116\pm0.042\pm0.004$ for Belle. For the direct \CP
asymmetry \babar measures
\begin{eqnarray}
A_{\CP}(B^0\rightarrow D^{*+}D^{*-})=-0.02\pm0.11\pm0.02, \nonumber
\end{eqnarray}
and Belle measures
\begin{eqnarray}
A_{\CP}(B^0\rightarrow D^{*+}D^{*-})=-0.16\pm0.13\pm0.02. \nonumber
\end{eqnarray}
Both measurements are consistent with each other and
consistent with no direct \CP violation. This favors in the
$B^0\rightarrow D^+D^-$ decays the \babar measurement to be
correct over the Belle measurement.

\subsection{$B^0\rightarrow \Jpsi\pi^0$}

The last discussed decay proceeding through the $b\rightarrow c\bar{c}d$
transition is the decay $B^0\rightarrow \Jpsi\pi^0$.
In this decay, the tree amplitude is Cabibbo- and
color-suppressed,
thus providing a useful laboratory to test the penguin
contribution from new physics beyond the standard model.

Both Belle and \babar analyzed this decay recently. The Belle
analysis is based on $535$ million $B\overline{B}$ pairs
which 
provides $290$ signal events \cite{Lee:2007wd}. The measured direct \CP
asymmetry is
\begin{eqnarray}
A_{\CP}(B^0\rightarrow \Jpsi\pi^0)=-0.08\pm0.16\pm0.05.
\nonumber
\end{eqnarray}
The analysis from \babar  is based on $466$ million
$B\overline{B}$ pairs which yields $184\pm15$ signal events
\cite{Aubert:2008bs}. The result for the direct \CP asymmetry is
\begin{eqnarray}
A_{\CP}(B^0\rightarrow \Jpsi\pi^0)=-0.20\pm0.19\pm0.03.
\nonumber
\end{eqnarray}
Measurements from both B-factory
experiments are consistent
with no direct \CP asymmetry as expected in the standard model. 

\subsection{Summary}

\begin{figure}[tb]
\begin{center}
 \includegraphics[width=7.0cm]{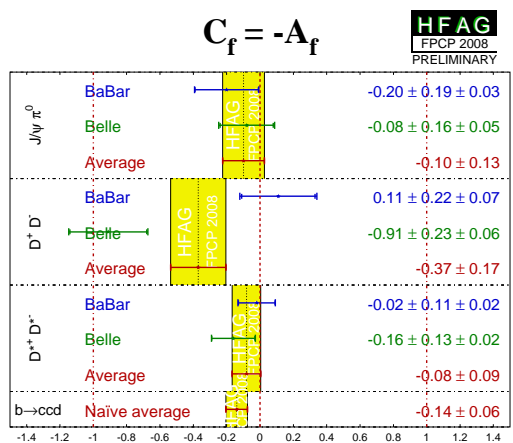}
 \caption{Summary of the current status of the direct \CP
violation in the $B^0$ decays governed by the $b\rightarrow
c\bar{c}d$ transition \cite{Barberio:2007cr}.
\label{fig12}}
\end{center} 
\end{figure}

A graphical summary of the recent results on the direct \CP violation in
the $B^0$ decays governed by the $b\rightarrow c\bar{c}d$
transition is shown in Fig.~\ref{fig12}.
Taking into account also $B^+$ decays governed by the same
transition, all measurements but $B^+\rightarrow
D^+\overline{D}^0$
from Belle  are consistent with no direct \CP
violation as expected by the standard model. This is an
indication
that in case of the $B^0\rightarrow D^+D^-$ decay the result observed
originally by Belle is most probably due to
a fluctuation rather than a sign of a new physics. On the other hand
averaging all measurements from the $B^0$ decays yields
\begin{eqnarray}
A_{\CP}(b\rightarrow c\bar{c})=-0.14\pm0.06, \nonumber
\end{eqnarray}
which is na\"\i vely $2.3$ standard deviations from zero.

\section{\BplusJpsiK decays}
\label{sec:5}

The last topic to discuss is a measurement of the direct \CP violation
in  $B^+\rightarrow \Jpsi K^+$ by D\O{} 
\cite{Abazov:2008gs}. This decay is governed by the $b\rightarrow
c\overline{c}s$ transition. In the standard model tree and
penguin amplitudes have a small relative weak phase.
Therefore one expects a small direct \CP violation on
a subpercent level. On the other hand new physics
contributions can enhance it to around 1\%
\cite{Hou:2006du,Wu:1999nc}. The measurement of the direct
\CP violation in this decay can be a clean way of observing
new physics, or important in constraining different models of new
physics. In addition the same analysis has access to the decay
$B^+\rightarrow \Jpsi \pi^+$ proceeding through
the $b\rightarrow c\overline{c}d$ transition which allows
another check of the large direct \CP asymmetry seen by
Belle  in the
$B^0\rightarrow D^+D^-$ decays.

\begin{figure}[tb]
\begin{center}
 \includegraphics[width=7.0cm]{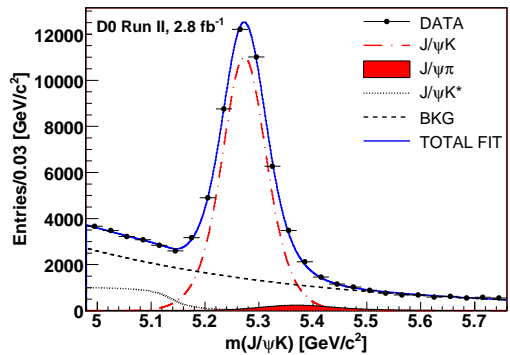}
\caption{Invariant mass distribution of the $B^+\rightarrow
\Jpsi K^+$ candidates from D\O{} 
\cite{Abazov:2008gs}. 
\label{fig13}}
\end{center} 
\end{figure}
The analysis uses $2.8$ \invfb of data collected by the D\O{}
detector using the dimuon trigger. The invariant mass distribution
of selected events is shown in Fig.~\ref{fig13}. After
selection, $40222\pm242$ $B^+\rightarrow \Jpsi K^+$ and
$1578\pm119$ $B^+\rightarrow \Jpsi \pi^+$ signal events are
available. The most important effect which needs to be
controlled is the asymmetry induced by the difference in the
interaction of kaons with the detector material. This
asymmetry is
measured directly from data using the decay $D^{*+}\rightarrow
D^0\pi^+$ with $D^0\rightarrow \mu^+\nu_\mu K^-$ and assuming no
direct \CP violation in the semileptonic $D^0$ decays. The asymmetry
due to the kaon interaction with material is found to be
$-0.0145\pm0.0010$. Taking this into  account, D\O{} measures
the direct \CP asymmetries
\begin{eqnarray}
A_{\CP}(B^+\rightarrow \Jpsi K^+)&=&0.0075\pm0.0061\pm0.0027,
\nonumber \\
A_{\CP}(B^+\rightarrow \Jpsi \pi^+)&=&-0.09\pm0.08\pm0.03.
\nonumber
\end{eqnarray}
Both asymmetries are consistent with zero as expected in
the standard model. The sensitivity of the asymmetry $A_{\CP}(B^+\rightarrow
\Jpsi K^+)$ is approaching the interesting region where it can
start to make constraints on new physics models.

\section{Conclusions}

The last year was very productive for experimental studies of
the direct \CP violation. We saw results from the \babar, Belle, CDF
and D\O{} experiments. 

The topic which attracted the most attention in the last year is the direct \CP
violation in the $B^0\rightarrow K^+\pi^-$ decay, which is most
precisely measured and the only one seen in a single experiment
with a significance of more than 5 standard deviations. The main point of
discussions is its difference from the decay $B^+\rightarrow
K^+\pi^0$, which could be a sign of new physics, but could
also be an effect of neglected standard model amplitudes, which could be
sizable. In connection to this result it will be interesting
to watch out for updates of the measurement in $B_s\rightarrow
K^-\pi^+$ decays. The current measurement by CDF is $2.5$ standard
deviations from zero and consistent with the large expected direct \CP
violation. With the next update we could have first
evidence for \CP violation in the $B_s$ system.

The second important achievement in the previous year is the first
analysis of the direct \CP violation in the charmless two body
$\Lambda_b$ decays from CDF. While its significance of being
nonzero is only $2.1$ standard deviations for the decay
$\Lambda_b\rightarrow pK^-$, the next update could result in
the first evidence of the direct \CP violation in the $b$-baryon sector.

The third important topic is the first evidence for direct \CP violation
in the $B^+$ sector in decays to $\rho K^+$. This is seen by
both Belle and \babar, with a chance that the asymmetry in the decay to $f_2^0(1280)
K^+$ will reach a significance of more than  $3$ standard deviations as well.

The last point which obtained considerable attention are
decays governed by the $b\rightarrow c\overline{c}d$ transition.
This interest was generated by the large \CP violation seen by
Belle in the $B^0\rightarrow D^+D^-$ decay. In the last year new results
most importantly on the $B^0\rightarrow D^{*+}D^{*-}$ and
$B^0\rightarrow \Jpsi\pi^0$ decays became available. For
both decays, measurements are consistent
with no \CP violation, but the na\"\i ve average of the
$B^0$ decays governed by the $b\rightarrow c\overline{c}d$
transition is $2.3$ standard deviations from zero.

Altogether the last year provided many exciting results in the area of
direct \CP violation with promises for the near future. In that we
might witness the first evidence for the direct \CP violation in
the $B_s$ system as well as in the $\Lambda_b$ system.

\begin{acknowledgments}
The author would like to thank all his colleagues from
the \babar, Belle, CDF and D\O{} experiments, who
contributed to the preparation of this talk and proceedings by
analyzing data, checking the material, and giving useful comments.
\end{acknowledgments}

\end{document}